\documentstyle[11pt,newpasp,twoside]{article}
\def\etal{{et~al.\ }}

\def\eg{e.g.,\ }
\def\subsun{_\odot}

\def\edcomment#1{\iffalse\marginpar{\raggedright\sl#1\/}\else\relax\fi}
\reversemarginpar

\begin{document}
\title{A Search for Intra-group Planetary Nebulae in the M81 Group}

\author{John J. Feldmeier}
\affil{Case Western Reserve University, 10900 Euclid Ave. 
Cleveland, OH 44106, U.S.A.}
\author{Patrick R. Durrell, Robin Ciardullo}
\affil{Penn State University, 525 Davey Lab, University Park, PA 16802, U.S.A.}
\author{George H. Jacoby}
\affil{WIYN Observatory, P.O. Box 26732, Tucson AZ 85726, U.S.A.}

\begin{abstract}

We present the initial results from an [O~III] $\lambda$ 5007 
survey for intra-group planetary nebulae in the M~81 group of 
galaxies.  A total of 0.36 square degrees of the survey have been 
analyzed thus far, and a total of four intra-group candidates 
have been detected.  These data allow us to probe the physics 
of galaxy interactions in small groups, and give us 
an upper limit for the density of intracluster starlight.
We find that the M~81 group has less than 3\% of its stars
in an intra-group component; this is much less than the fraction 
seen in richer galaxy clusters.  
\end{abstract}

Intracluster planetary nebulae (IPN) trace  
intracluster starlight, and provide crucial information on the 
kinematics of intracluster stars.  However, although the presence 
of IPN in massive clusters such as Virgo and Fornax has now 
been well established (Feldmeier, this conference), the amount of 
intracluster planetaries in much poorer environments is unknown.  
Searching poorer systems for IPN is a promising, but currently 
unexplored, avenue of research.  If most intracluster stars are removed by 
galaxy collisions (\eg Richstone \& Malumuth 1983; 
Moore \etal 1996), the fraction of intra-group stars 
will be significantly less than that seen in a rich cluster.

As a first step, we have begun a large-scale [O~III] $\lambda$5007 IPN 
survey of the nearby M~81 group of galaxies.  This galaxy group is 
known to be strongly interacting, with multiple tidal tails seen 
in H~I gas (Yun, Ho, \& Lo 1994).  Using the KPNO 4-meter telescope and
the MOSAIC imager, we have observed 0.72 square degrees of the group, 
and we have analyzed the first 0.36 square degrees.  Using our 
automated detection methods (\eg Theuns \& Warren 1997; 
Feldmeier 2000), we found a total of 22 PN candidates in our 
first field, which can be divided into three components:
\pagebreak
\begin{itemize}

\item 16 PN candidates surrounding the elliptical galaxy NGC~3077, which
is in the southeastern corner of the field.

\item 2 PN that can be attributed to M~81's (NGC~3031) outer 
halo, which is directly west of this field.

\item 4 intra-group PN candidates.

\end{itemize}

Of these four intra-group candidates, only a single object 
is above the photometric
completeness level.  If we assume this candidate is bona-fide, 
we can set an upper limit for the intra-group luminosity.  In order 
to be conservative, we adopt a generous conversion factor
between PN density and luminosity ($\alpha_{2.5} = 16 
\times 10^{-9} L\subsun$), the amount derived for the bulge of 
M~81 (Jacoby \etal 1989).  We find that, at most, there is 
$\sim 2 \times 10^{8} L\subsun$ of intra-group stars in our Mosaic 
field.  When we compare this number to the total luminosity and angular 
area of the M~81 group (Garcia 1993), we find that less than 3\% of the 
group's stellar luminosity is intra-group in nature.  This is much 
smaller than the $\sim 20\%$ found for the Virgo and Fornax clusters 
(Feldmeier 2000; Durrell \etal 2001; Feldmeier \etal 2002).  This 
result is consistent with tidal stripping scenarios.

Our PN survey is ongoing, and we will be able to place much more 
stringent limits on the intra-group luminosity surface density.  
We have also begun a broad-band survey for intra-group red giant 
stars in the M~81 group using the {\sl CFHT}.  The two methods 
complement each other: red giant surveys allow 
an independent check of our PN survey, and can probe to fainter 
surface brightnesses.  On the other hand, through follow-up spectroscopy, 
IPN can provide dynamical information and nebular abundances.
Finally, we also plan an ultra-deep imaging survey of the M81 group 
using the Burrell Schmidt.  We will perform high-quality 
surface photometry to obtain the global picture of this system.

\end{document}